\documentclass{article}

\usepackage{PRIMEarxiv}

\usepackage[utf8]{inputenc} 
\usepackage[T1]{fontenc}    
\usepackage{hyperref}       
\usepackage{url}            
\usepackage{booktabs}       
\usepackage{amsfonts}       
\usepackage{nicefrac}       
\usepackage{microtype}      
\usepackage{lipsum}
\usepackage{fancyhdr}       
\usepackage{graphicx}       
\usepackage[dvipsnames]{xcolor}
\graphicspath{{media/}}     

\title{Thermoelectric Properties of Graphene through BN-ring Doping: A Theoretical Investigation}

\author{  Laura Caputo, Viet-Hung Nguyen, Jean-Christophe Charlier \\
  Institute of Condensed Matter and Nanosciences, \\
  Université catholique de Louvain (UCLouvain),\\
  Chemin des étoiles 8, B-1348, Louvain-la-Neuve, Belgium\\
}

\begin{document}
\maketitle

\begin{abstract}
Graphene has been widely studied for various applications due to its outstanding electrical and mechanical properties. However, its potential in thermoelectric applications has been limited by a low Seebeck coefficient and high thermal conductivity. Efforts to enhance its thermoelectric properties have involved the usage of carbon-based nanoribbons \cite{Zheng2012,Hossain2016}, strain engineering \cite{Nguyen2015}, and heteroatom co-doping, particularly with Nitrogen and/or Boron atoms \cite{Wang2018}. In this work, multiscale simulation approaches combining DFT calculations and semi-empirical models are used to explore the potential improvement of the thermoelectric properties via borazine (B$_3$N$_3$)-ring doping. As bandgap engineering can be obtained with this doping, the thermoelectric properties of graphene are significantly enlarged, albeit at the cost of reduced conductance. The effects observed are not only dependent on the concentration of BN within the graphene lattice but are also notably influenced by the relative rotational alignment of the BN rings. Furthermore, the effect of the distribution and rotational disorder are considered, showing reduced electronic conductance compared to the periodic case highlighting the importance of precise control over the doping parameters of BN-ring. Lastly, the thermal lattice conductance is estimated revealing a substantial reduction of up to 40$\%$ compared to pristine graphene opening up the possibility of enhancing the thermoelectric efficiency of BNC materials. The present theoretical approach highlights how BN-ring doping can refine the thermoelectric properties of 2D graphene, offering a pathway for enhancing its suitability in practical thermoelectric applications.
\end{abstract}


\section{Introduction}
Despite the thermoelectric effect has been known for nearly two centuries, its applications continue to expand, making thermoelectric materials a crucial focus of research and development in the pursuit of sustainable and efficient energy solutions. In particular, the attractiveness of the thermoelectric effect lies in its potential to harness waste heat and convert it into valuable electrical power. This renders thermoelectric materials highly relevant in various applications, including automotive waste heat recovery \cite{Orr2016}, spacecraft power generation \cite{Ritz2004}, and wearable energy harvesting devices \cite{Nozari2020}. While there is significant interest in flexible thermoelectric materials for portable and wearable electronics, conventional inorganic bulk materials remain rigid and fragile, rendering them impractical for device applications. Substantial efforts have been therefore dedicated to discovering alternative materials capable of replacing current thermoelectric ones. Since the groundbreaking research by Hicks and Dresselhaus \cite{Hicks1993}, the focus on nanostructuring approaches to form low-dimensional systems has emerged as a dominant strategy to bolster thermoelectric properties \cite{Jia2019,Olaya2017}. Their research indicates that enhancing the energy-dependent nature of conductivity ($\sigma$) could notably augment the Seebeck coefficient ($\textit{S}$), i.e. the thermoelectric sensitivity of the conductor. Consequently, low-dimensional systems are anticipated to offer a higher Seebeck coefficient and accordingly higher power factor (\textit{PF} = $S^2\sigma$), compared to bulk materials.
Note that the thermoelectric efficiency is quantified by the figure of merit $ZT = \sigma S^2 T / \kappa$, where $\kappa$ represents the thermal conductivity. Therefore, two well-known strategies for improving $ZT$ are either to enhance the power factor or to reduce the thermal conductivity. The enhancement of the power factor serves as an additional advantage by contributing to the improvement of the output power of thermoelectric generators \cite{Liu2016}. Building on this foundation, graphene stands out as a promising candidate for thermoelectric applications due to its exceptionally high electrical conductivity $\sigma$. 

Indeed, electrons mimic the electronic properties of massless Dirac fermions in the vicinity of the high-symmetry K-points of the Brillouin zone of graphene, leading to particularly high energy-independent group velocity ($\sim$ 10$^6$cm s$^{-1}$) \cite{Castro2009}. Despite its high electric mobility and high electrical conductivity, graphene faces notable challenges when considering its application in thermoelectric devices. In fact, its semi-metallic nature, presenting challenges in discerning opposite contributions from electrons and holes, leads to a low Seebeck coefficient \cite{Seol2010}. Nevertheless, despite these intrinsic obstacles, notable efforts have been invested in improving the thermoelectric properties of graphene. First and foremost, published studies have focused on the engineering of the Seebeck effect in graphene, particularly, opening a bandgap in order to separate the contributions of electrons and holes \cite{Dollfus2015}. Striking examples include graphene nanoribbons \cite{Mazzamuto2011,Sevincli2013,Tran2015,Tran2017,Chang2014,Deng2019}, graphene antidot superlattices \cite{Karamitaheri2011,Gunst2011,Jinwoo2O17,Hung2014} and strain engineering \cite{Nguyen2015}. Nanostructuring techniques to improve the thermoelectric figure of merit ZT have been also reported \cite{Mazzamuto2011,Sevincli2013,Chang2014,Hung2014,Deng2019}.

A two-dimensional material that bears a striking resemblance to graphene in terms of its structure and electron configuration is hexagonal boron nitride (\textit{h}-BN). In fact, both materials are isostructural and isoelectronic. Similar to graphene, \textit{h}-BN is characterized by a honeycomb lattice can be described in terms of two interlocking sublattices. However, the presence of two different types of atoms within the sub-lattices of pristine \textit{h}-BN disrupts its inversion symmetry, which allows for lifting the degeneracy at the Dirac points within the Brillouin zone. This results in a substantial band gap of approximately 6 eV in \textit{h}-BN \cite{Cassabois2016} which inherently limits its application as a semiconductor in electronics, as this wide band gap restricts electron mobility under normal conditions \cite{Wong2021}. Thereafter, introducing BN, particularly in structures such as borazine-like rings, into graphene allows to strategically break its symmetrical structure. This sublattice symmetry breaking is a key to engineering a band gap in graphene gaining the ability to tailor its electronic properties opening up new possibilities for creating graphene-based electronic devices with customizable functionalities.
Therefore, considerable efforts have been dedicated to discovering novel synthetic methodologies for the successful production of periodic BN-doped covalent carbon networks, ensuring consistent doping patterns. In fact, the successful synthesis of a BN-doped covalent network with a periodic doping pattern has been accomplished by using bottom-up techniques \cite{Sanchez2015}. However, this represents the only example of successful synthesis, leaving the properties and potential applications of these BNC monolayers yet to be fully explored.
Therefore, theoretical studies serve as a crucial methodology to facilitate the tailored design of new synthetic approaches aimed at producing specific BNC monolayers with desired properties. In fact, extensive theoretical investigations into the effect of Boron or Nitrogen doping effect on graphene thermoelectric properties through single-atom doping \cite{Mann2020,Abdullah2020,Mortazavi2012} or \textit{h}-BN flakes \cite{Sevinccli2011,Tran2015,Yang2012,Dong2024} have been reported. However, to the best of our knowledge, the impact of isolated borazine ring doping on the thermoelectric properties of graphene remains unexplored. The impact of borazine ring doping not only offers a promising avenue to manipulate the electronic characteristics of graphene but it has also been shown to have a significant dependence on various doping parameters, including doping concentration, ring rotation, and ring position \cite{Caputo2022}.

As mentioned, one drawback of graphene-based materials in thermoelectric applications is the co-contribution of electrons and holes, leading to a weak Seebeck effect in pristine graphene materials. The above-discussed BN-ring doping could be alternatively a promising technique to improve the thermoelectric properties of graphene. In such context, the current work reports multi-scale calculations of the electronic thermoelectric properties of BN-ring doped 2D-graphene monolayers, emphasizing the influence of different doping parameters, including borazine ring concentration and orientation. First, \textit{ab}-initio calculations are performed for systems obtained when BN-rings are periodically inserted into graphene. Non-periodic doped systems are then computed using a semi-empirical tight-binding model to depict more realistic scenarios and highlight the impact of the disorder on these properties. At last, to estimate the thermoelectric efficiency of the considered BNC materials, the lattice thermal conductivity and accordingly the figure of merit ZT have been calculated using semi-empirical approaches.

\section{Methologies}
\begin{figure}[ht]
    \centering
    \includegraphics[width=13cm]{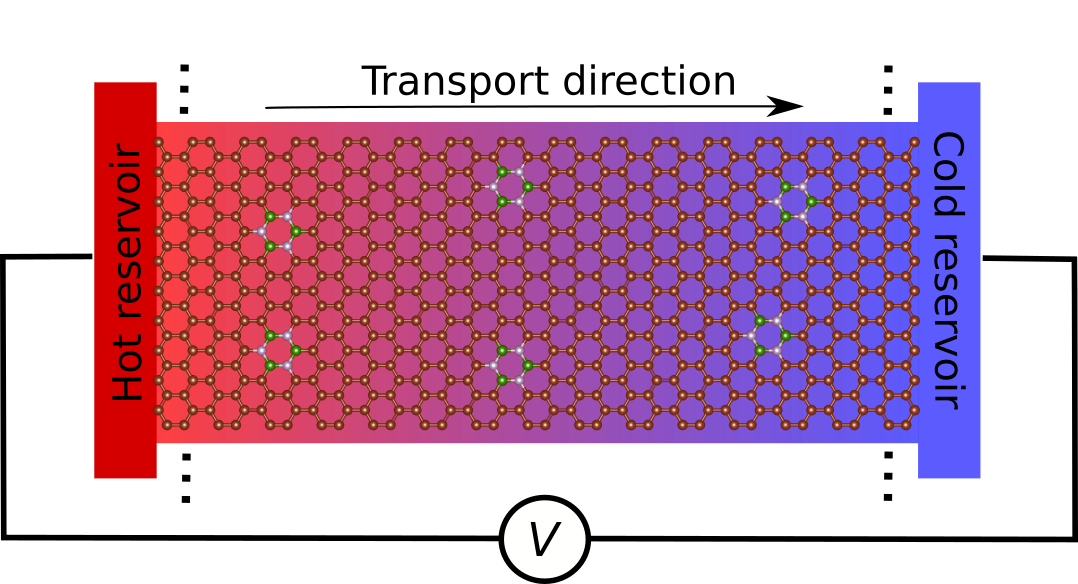}
    \caption{Schematic model of a typical thermoelectric device built with a BNC material. The transport direction is defined from the hot (red) to the cold (blue) reservoir. The central region is formed by a periodically Boron and Nitride co-doped graphene layer. The model is periodic in the direction perpendicular to the transport, denoted with dashed black lines. Brown, green and grey spheres represent Carbon, Boron and Nitrogen atoms, respectively.}
    \label{fig:device}
\end{figure}

Transport calculations have been performed on a device, as schematized in Figure \ref{fig:device}. In particular, a BNC monolayer is considered as the device channel transmitting charge carriers and heat between left (hot) and right (cold) contacts (reservoirs). The channel is assumed to be very large in the perpendicular direction and hence can be computed using the periodic boundary condition in such direction.

As discussed, periodically doped BNC monolayer can be achieved using bottom-up techniques \cite{Sanchez2015}. Therefore, our study first focuses on such cases and first-principles calculations are conducted on thermoelectric properties of periodically-doped graphene-based materials. These calculations are performed using the combination of Density Functional Theory (DFT) framework and Green's function (GF) approach in the ballistic approximation, as implemented in OpenMX code \cite{Ozaki2003,Ozaki2004,Ozaki2005}. The generalized-gradient approximation of Perdew Burke and Ernzherof (PBE) \cite{Perdew1996} has been exploited for the exchange-correlation density functional and the electron-ion interaction is described with norm-conserving pseudopotentials as proposed by Morrison et al. \cite{Morrison1993}. $\textit{Ab initio}$ calculations have been performed with the usage of a numerical pseudo-atomic orbitals as basis set, decomposed in two \textit{s}-orbitals, two \textit{p}-orbitals and one \textit{d}-orbital for each of the atom types. The energy cutoff for real-space mesh has been chosen to be 300Ry while a k-point grid of 33x33x1 for the graphene unit cell has been employed. Moreover, an interlayer vacuum distance of 20\AA{} has been employed to avoid spurious interaction along the non-periodic $\textit{z}$ direction. To investigate different BN doping scenarios, a 5x5 supercell was built from the orthorhombic unit cell of graphene and consequently doped with BN rings to create models with varying concentrations. Each atomistic model underwent full relaxation of in-plane cell parameters and atomic positions.
The possibility of periodically synthesizing BNC monolayers using bottom-up techniques, and thus tailored precursors, opens up the possibility of potentially controlling the orientation of the BN rings as well. The latter, in fact, can strongly affect the electronic properties of BNC materials \cite{Caputo2022}. Specifically, two distinct scenarios arise: one where all the rings possess the same orientation (referred to as the parallel orientation case) and another where a mixture of BN rings exhibits opposite orientations (referred to as the anti-parallel orientation case). In our \textit{ab initio} calculations, both the parallel and anti-parallel cases are explored, with the latter examined across various concentrations of anti-parallel rings whenever feasible.
Furthermore, the thermoelectric parameters were derived from electronic transmission ($T_e$) using the Landauer-B{\"u}ttiker formalism. Specifically, this involves defining the auxiliary function $L_n$ as
\begin{equation}
    L_n(\mu,T) = \int T_e(E) (E-\mu)^n \left[ -\frac{\partial  f_e(E,\mu,T)}{\partial E} \right] dE
\end{equation}
where $\mu$ is the electron chemical potential and $f_e (E,\mu,T)$ is the Fermi-Dirac distribution function. The electronic conductance $\textit{G}$, the Seebeck coefficient $\textit{S}$ and the electronic thermal conductance $\textit{K}_e$ can then be calculated as
\begin{equation}
    G(\mu,T) = \frac{2e^2}{h}L_0(\mu,T)
\end{equation}
\begin{equation}
    S(\mu,T) = \frac{1}{eT}\frac{L_1(\mu,T)}{L_0(\mu,T)}
\end{equation} 
\begin{equation} 
    K_e(\mu,T) = \frac{1}{T} \left[ L_2(\mu,T)-\frac{L_1(\mu,T)^2}{L_0(\mu,T)} \right]
\end{equation}

The dynamical properties of these atomic models were computed utilizing the fourth nearest-neighbor force constant (4NN FC) model. To this aim, the FC parameters were taken from \cite{Wirtz2004} for C-C bonds and from \cite{Xiao2004} for B-N bonds. For the coupling of C atoms with B and N atoms, the average values of the FC parameters between graphene and $\textit{h}$-BN have been considered \cite{Yang2012,Tran2015}.
Lastly, the GF method has been employed to compute the phonon transmission $T_p$, in which the surface GF $G_{L(R)}^0$ of the leads is computed utilizing the Sancho-Rubio scheme \cite{LopezSancho1984}.
Consequently, the lattice thermal conductivity has been calculated as
\begin{equation}
    K_l = \int_0^{\infty} \frac{d\omega}{2\pi} \hbar\omega T_p(\omega)\frac{\partial  n(\omega,T)}{\partial  T} 
\end{equation}
where, the Bose-Einstein distribution function $n(\omega,T)$ is now considered. To determine the efficiency of each BNC model, the thermoelectric figure of merit has been evaluated using the standard formula : 
\begin{equation} \label{eq:ZT}
    ZT = \frac{S^2G}{K_e+K_l} \, T
\end{equation}
Note that here, the total thermal conductance is a sum of electron contribution ($K_e$) and lattice thermal part ($K_l$).

Finally, to complete the study, aperiodic systems are also investigated.
As the size of computed systems is large, DFT calculations are therefore unfeasible and the electronic properties were considered using a tight-binding (TB) model \cite{Caputo2022} (see section 3 of Supplementary Material). The methodology mirrors the one described previously for calculating the dynamical matrix and consequently the lattice thermal conductivity, with the only difference that in this case the electronic Hamiltonian is considered to calculate the thermoelectric properties. Furthermore, owing to the considerably larger size of the models required to accommodate non-periodic doping, the computation of the Green's function was executed using a recursive technique \cite{Nguyen2023} that permits to compute recursively the Green's function blocks.
In addition, note that the electron-phonon interactions have been shown to be notably weak in monolayer graphene \cite{Du2008}, which have been experimentally demonstrated by high carrier mobilities \cite{Geim2007,Yankowitz2019,Pulizzi2019} and micrometer-scale ballistic transport length \cite{Baringhaus2014,Barrier2020}. Therefore, electron-phonon interactions are approximately neglected in our transport calculations.
Throughout the work, calculations were performed at room temperature.

\section{Discussion}
\subsection{Improved thermoelectric effects}
\begin{figure}[!ht]
    \centering
    \includegraphics[width=15.7cm]{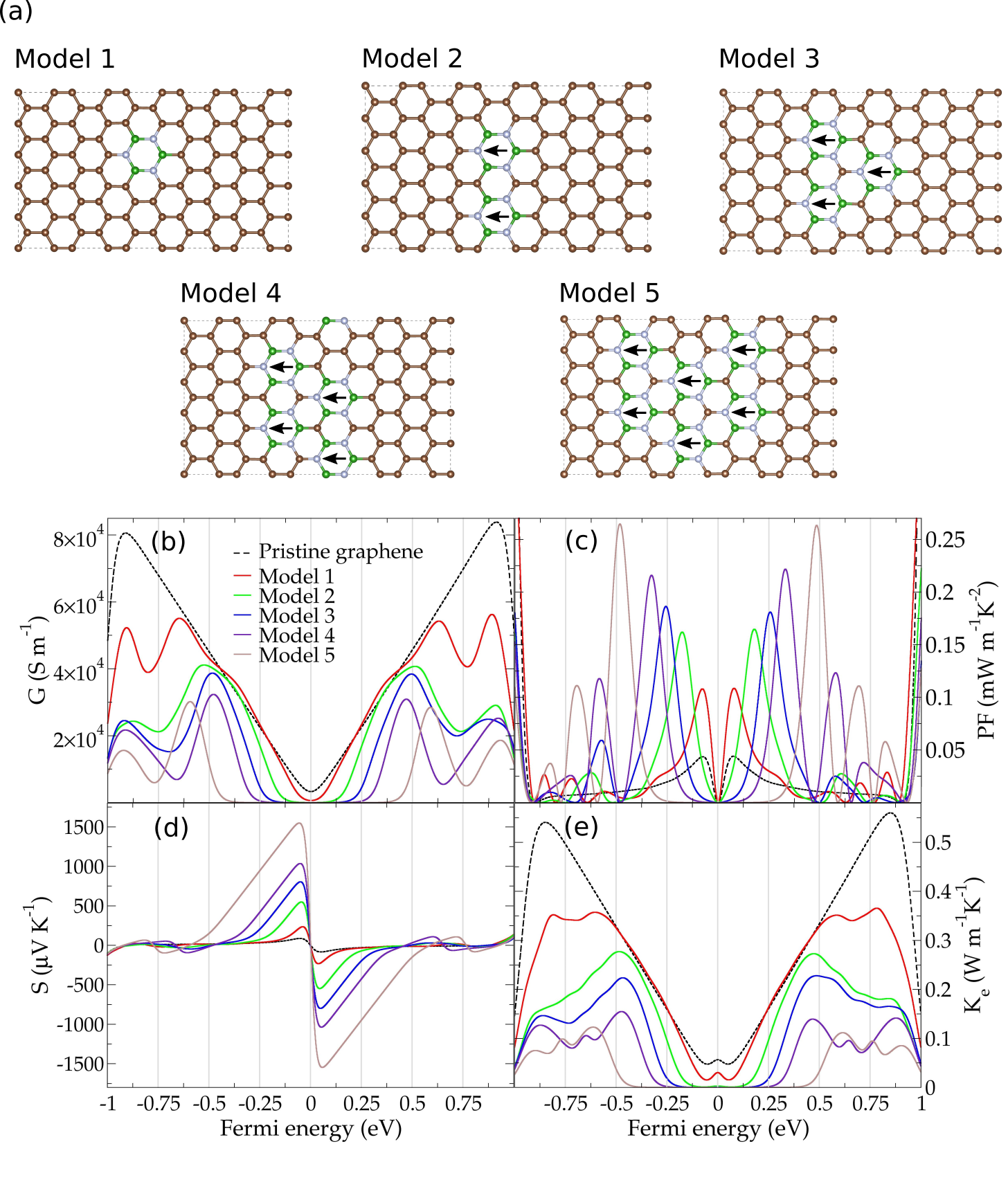}
    \caption{(a) Atomistic models of graphene doped with various BN-ring concentrations in the parallel orientation and their corresponding (b) electronic conductance (\textit{G}), (c) power factor (\textit{PF}), (d) Seebeck coefficient (\textit{S}), (e) electronic thermal conductance ($K_e$) as a function of the Fermi level. Pristine graphene is reported with a black dashed line as a reference.}
    \label{fig:upup}
\end{figure}

In evaluating the impact of borazine-like ring concentration and orientation on the thermoelectric properties of monolayer graphene, the effect of BN concentration in monolayer graphene is first considered. To this aim, six distinct percentage concentrations (Model 1 - 6$\%$, Model 2 - 12$\%$, Model 3 - 18$\%$, Model 4 - 24$\%$ and Model 5 - 36$\%$) were examined by progressively introducing an increasing number of borazine rings into the graphene supercell (two, three, four and six rings, respectively). Specifically, to highlight the effect of the concentration alone, the orientation of the BN-rings has been firstly kept fixed to the parallel orientation for each concentration (Fig. \ref{fig:upup}a). The electronic conductances ($G$), Seebeck coefficients ($\textit{S}$), Power Factors ($PF=S^2G$) and electronic thermal conductances ($K_e$) for each concentration model are reported in Figure \ref{fig:upup}.
As mentioned above, pristine graphene exhibits ambipolar characteristics (i.e., semi-metallic behavior) that allow both electrons and holes to present significant contributions to its thermoelectric properties when the charge-carrier concentration is low. Therefore, the opposite contributions of these two carrier types lead to a small Seebeck effect. In particular, a Seebeck coefficient smaller than 100$\mu$VK$^{-1}$ at room temperature has been experimentally reported \cite{Zuev2009}. Indeed, this aligns closely with our calculations, as presented in Fig.\ref{fig:upup}d.

When BN doping is implemented, the Seebeck coefficient and the PF significantly increase, achieving values of more than an order of magnitude higher than that of pristine graphene. In particular, for both properties, a nearly linear trend with the BN concentration is found. Note that enhancing the PF is beneficial for improving the output power density, $\omega$, of a thermoelectric generator \cite{Liu2016}. The output power density is a measure of how much power per unit volume (or mass) the generator can produce, and it is directly proportional to the PF, given by the equation $\omega = PF \cdot \Delta T^2/4L$, where $\Delta T$ is the temperature difference across the thermoelectric material and \textit{L} is the thermoelectric leg length.
Therefore, improvements in the PF directly translate to higher output power densities, making devices more compact and efficient for a given power output. However, this enhancement comes at the cost of reduced electrical conductivity and band gap opening. In particular, when introducing BN-rings in the parallel orientation, the sublattice symmetry of graphene is broken due to the strongly different composition of each sublattice. In fact, in this particular orientation configuration, each Nitrogen (Boron) dopant atom occupies the position of a Carbon atom in the A (B) sublattice, resulting in the strongest asymmetrical state for a given concentration. Consequently, this perturbation leads to the opening of a band gap in graphene \cite{Caputo2022}, as shown in the conductance plot in Fig.2b. This effect, even if it strongly reduces the electronic conductance, allows the separation of the electron and hole contributions leading to a strong increase of the Seebeck coefficients compared to the semi-metallic case of pristine graphene. Nevertheless, the conductance data reveals an anticipated trend of band gap widening as the BN ring concentration increases. Note that at room temperature, the thermal broadening width measures around 0.2eV. A band gap of approximately this value would allow for the effective separation of electron and hole contributions, hence enhancing the Seebeck coefficient, while still maintaining a high electronic conductance at room temperature. This compromise would be ideal for practical thermoelectric devices. In particular, for 12$\%$ BN concentration corresponding to model 2, a band gap comparable to the desired value emerges.
\begin{figure}[!ht]
    \centering
    \includegraphics[width=15.7cm]{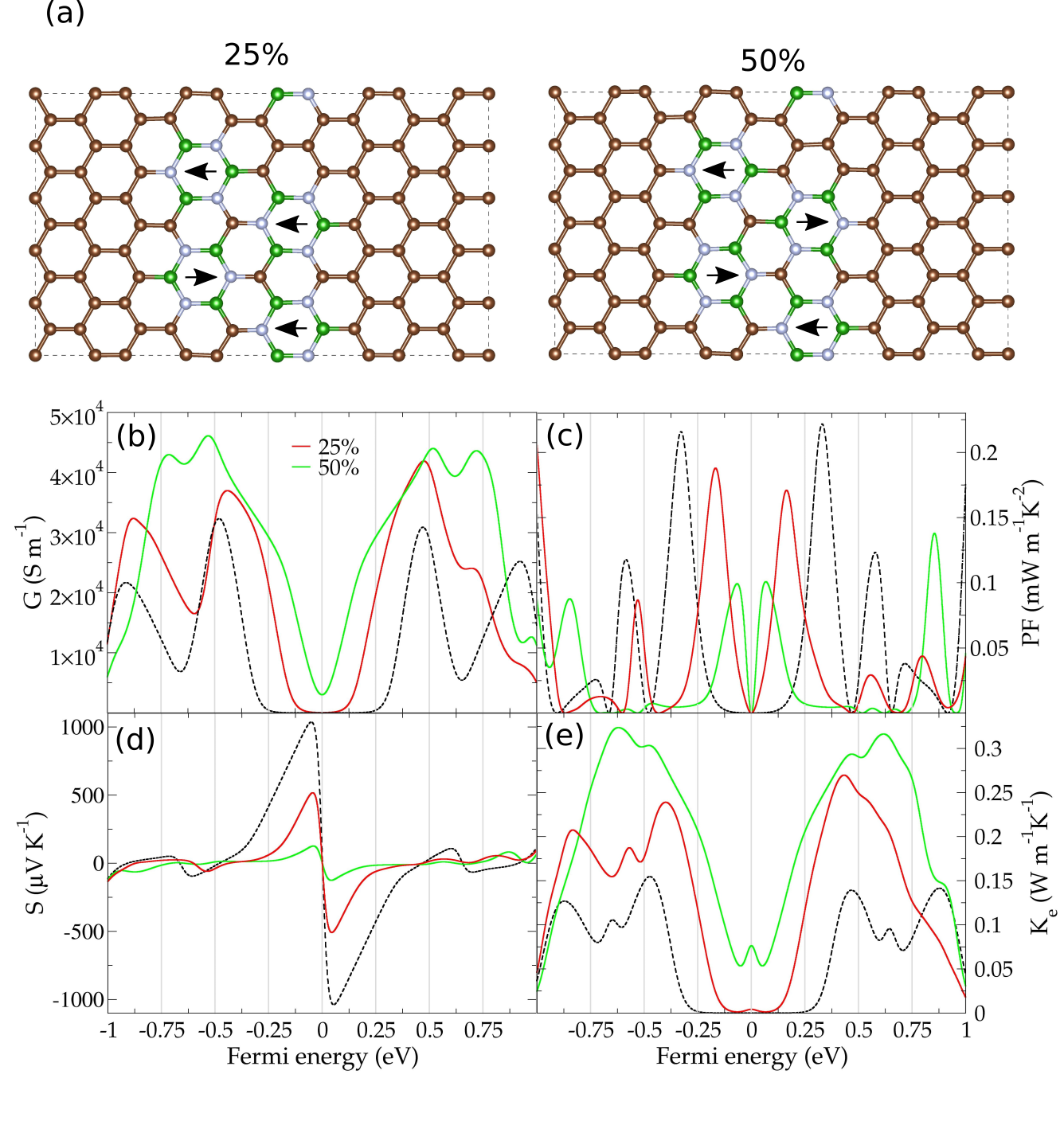}
    \caption{(a) Atomic structures of Model 4 with 25\% and 50\% anti-parallel rings and their corresponding (b) electronic conductance, (c) power factor, (d) Seebeck coefficient and (e) electronic thermal conductance. The same Model 4 with all parallel orientation of the BN-rings pointing in the same direction is reported as a black dashed line.}
    \label{fig:model4}
\end{figure}

Concerning the electronic contribution to the thermal conductance $K_e$, an effective bandgap emerges with BN concentration exceeding 12$\%$, while the lowest concentration investigated follows the trend of pristine graphene up to values of the Fermi energy of 0.5eV. While maximizing electronic conductance is essential for achieving high PF values, minimizing the electronic thermal conductance is crucial to enhance thermoelectric efficiency \textit{ZT} (Eq. \ref{eq:ZT}). 
Notably, introducing BN-ring doping in graphene strongly reduces the electronic thermal conductance, potentially enhancing its thermoelectric properties. However, as detailed in section 3.3, the $K_e$ values are lower than the lattice thermal conductance values, aligning with expectations and underlying the importance of the latter in the assessment of the thermoelectric efficiency.
The dependence on the distribution of BN-rings within the DFT supercell (see an example, Model 2, presented in section 1 of Supplementary Material) is also examined to generalise the results obtained until now. In particular, the transport properties around the bandgap and accordingly thermoelectric effects are shown to be weakly affected when changing the arrangement of BN rings.

Another important doping parameter that can be controlled to tune the thermoelectric properties is the orientation of the BN-rings. So far, BN-rings oriented in the same direction, establishing a parallel orientation configuration, have been considered. However, anti-parallel orientation configurations, i.e. rings having opposite orientations, can be also modelled within a periodic framework including multiple BN-rings in the unit cell. Consequently, the possibility of tuning the thermoelectric properties with the orientation of the rings has been considered here in models containing two or more BN-rings.
In simple cases involving only two BN rings, the parallel scenario entails all BN-rings having the same orientation, while the anti-parallel scenario involves 50\% rings oriented in each direction. Yet, when the graphene supercell in DFT calculations includes more than two BN rings, the number of rings oriented in each direction can vary. In these instances, the percentage of anti-parallel rings is introduced and defined as the ratio between the number of BN rings oriented in opposite directions. Examining models with high BN-ring concentrations enables the evaluation of a broader range of rotated ring percentage. Consequently, starting from Models 4 and 5 in Fig.\ref{fig:upup}a where all the rings share the same orientation (parallel case), a specific number of BN rings were rotated to create various anti-parallel models featuring different percentages of anti-parallel BN-rings (Figs. \ref{fig:model4},\ref{fig:model6}). The similar investigation of Models 2 and 3 is also presented in section 2 of Supplementary Material.
Despite thermoelectric properties being sensitive to the percentage of anti-parallel BN-rings, their transport properties around the bandgap and accordingly thermoelectric effects are not significantly affected by the exact positioning of these rotated rings within the DFT supercell, leading us to prioritize configurations with the representative properties for each level of rotated ring percentage.

\begin{figure}[!ht]
    \centering
    \includegraphics[width=15.7cm]{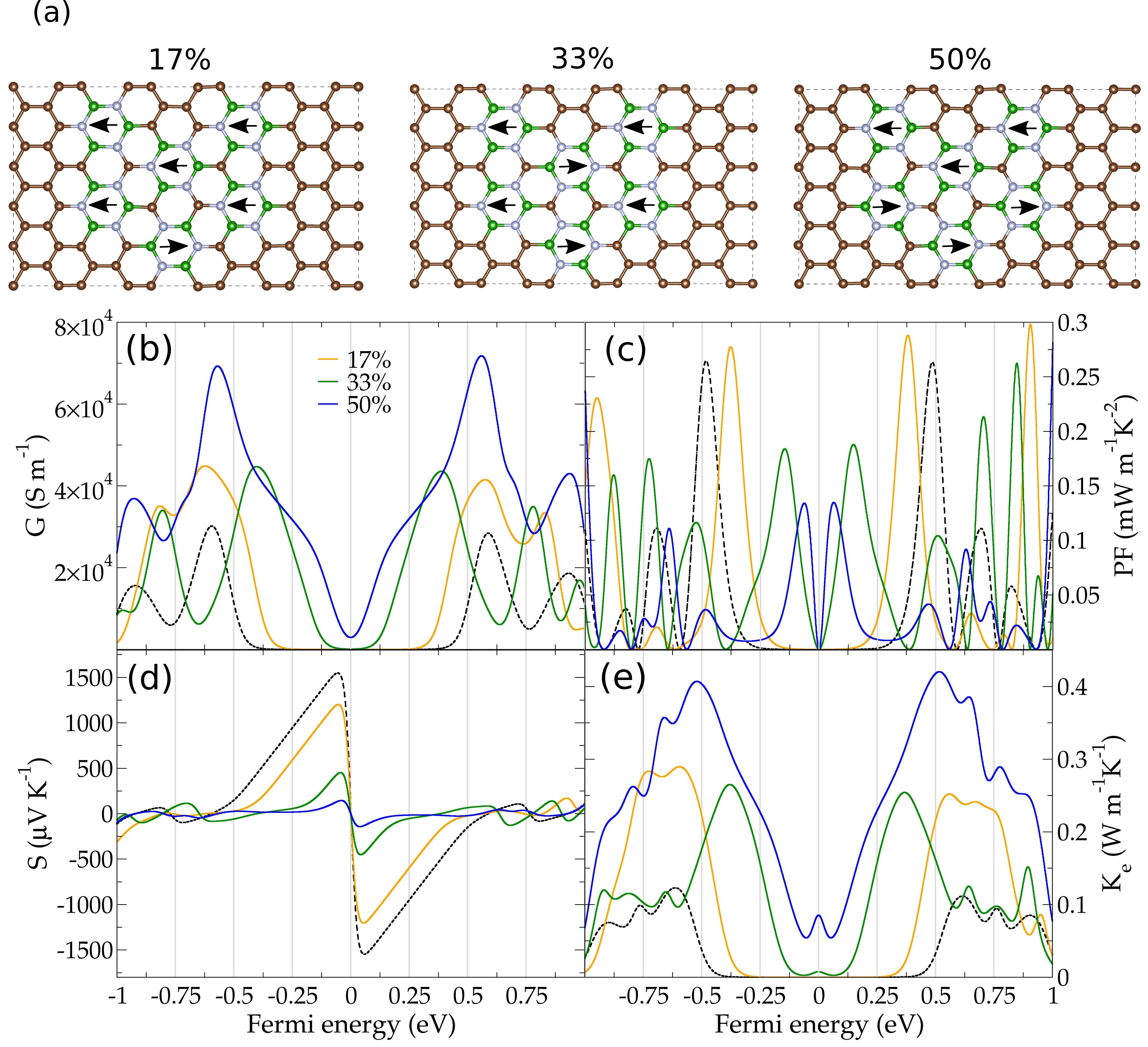}
    \caption{(a) Atomic structures of Model 5 with 17\%, 33\% and 50\% anti-parallel rings and their corresponding (b) electronic conductance, (c) power factor, (d) Seebeck coefficient, (e) electronic thermal conductance. The same Model 5 with all parallel orientation of the BN-rings pointing in the same direction is reported as a black dashed line.}
    \label{fig:model6}
\end{figure}

In both Model 4 (Fig. \ref{fig:model4}) and 5 (Fig. \ref{fig:model6}), combining the effect of the high BN-ring concentration with the fine-tuning of the anti-parallel ring percentage (25$\%$ for Model 4 and 33$\%$ for Model 5), band gaps close to 0.2eV can be attained as well as an enhanced conductance compared to the correspondent parallel scenario. Furthermore, the progressive restoration of sublattice symmetry becomes evident with the gradual increase in the percentage of anti-parallel rings, ultimately leading to the closure of the band gap. This phenomenon is observable in the Seebeck coefficients, where an effective separation of electron and hole contributions occurs at small percentages of anti-parallel rings. In the 50$\%$ anti-parallel ring cases, values resembling those of pristine graphene, indicating semi-metallic behaviour, are obtained. Similarly to Models 2 and 3, around the gap region values of PF close to the correspondent parallel models can be achieved with a small percentage of rotated rings, reported in dashed black lines. Lastly, the electronic thermal conductance aligns with the trend of the electronic conductance. However, as mentioned earlier, the enhancement of the thermal conductance, compared to the correspondent parallel orientation case, can be detrimental to the improvement of thermoelectric efficiency. In general, this emphasizes the effects of the chiral symmetry breaking (induced by the presence of BN rings as discussed in detail in Ref.\cite{Caputo2022}) underscoring its substantial impact on the thermoelectric properties of doped graphene. Indeed, by strategically combining the two doping parameters of BN-ring concentration and rotation, one can finely adjust the thermoelectric properties of BNC materials. This allows for the attainment of an optimal equilibrium between conductance and the Seebeck coefficient, potentially enhancing the power factor.

\subsection{Disorder effects}
\begin{figure}[!ht]
    \centering
    \includegraphics[width=15.7cm]{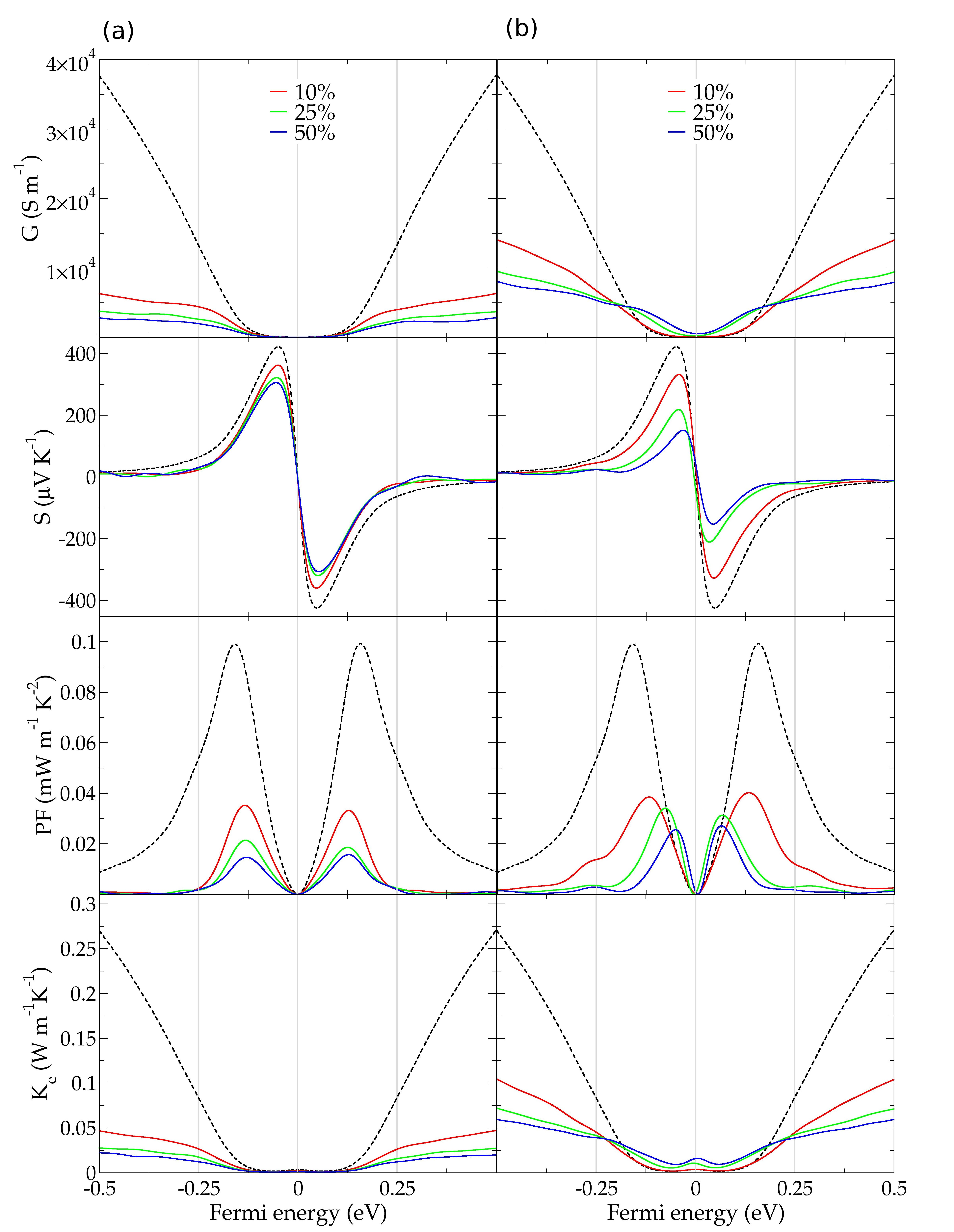}
    \caption{Electronic conductance, Seebeck coefficient, Power Factor and electronic thermal conductance for (a) distribution disorder and (b) rotational disorder computed within a tight-binding approach. Different disorder strengths 10\%, 25\% and 50\% are investigated while the perfectly periodic case is also reported in dashed black line. The BN-ring concentration is 12.5\% in all cases.}
    \label{fig:TB}
\end{figure}

The current advancements in the targeted synthesis of BN-doped 2D materials with specific and controlled properties have yet to reach the desired goal. In fact, BN-doped carbon-based materials lack periodic doping, resulting in properties that are nonreproducible \cite{Ci2010,Herrera2021}. In particular, creating materials with precisely controlled doping patterns of B and N atoms presents significant challenges due to atom segregation \cite{Huang2015,Yuge2009} arising from the strong binding energy between boron-nitrogen and carbon-carbon atoms.
Up to this point, the investigation has primarily centered on periodically doped BNC monolayers, being the main target in bottom-up synthetic methodologies. Nonetheless, attaining absolute control over the periodicity of the system still presents practical uncertainties. Consequently, tight-binding (TB) calculations have been performed to explore the impact of weak disorder on the electron thermoelectric properties. In particular, to avoid the sensitivity of the obtained results to the specific disorder configuration, a large model $\sim$154nm long and $\sim$37nm wide was built starting from a 4x3 supercell of orthorhombic graphene containing a single BN-ring and connected to pristine graphene electrodes. Consequently, from these large periodic models, the effect of different kinds of disorders has been studied. In particular, two types of disorder are considered: distribution disorder and rotational disorder (similarly, see the details of these disorder models in \cite{Caputo2022}). In the former case, a certain number of BN rings has been shifted from their original positions in the ideally structured system to adjacent locations. In particular, the focus lies on small displacements due to the main goal of achieving periodicity in experimental bottom-up samples. On the other hand, the rotational disorder occurs when a subset of BN rings within systems of parallel rings is rotated to an anti-parallel orientation relative to the original position of the ring in the periodic system, similar to the rotational disorder explored in the \textit{ab initio} calculations. Varying concentrations of disorder were explored for each type. Figure \ref{fig:TB}a illustrates the impact of the distribution disorder,  while Figure \ref{fig:TB}b showcases the effect of diverse rotational disorder strengths. In both cases, the perfectly periodic system computed using the same TB methodology is reported in dashed black lines.

These results align with previous studies \cite{Caputo2022}, where tight-binding calculations are shown to mimic DFT calculations thus showing the independence of the band gap with the distribution disorder, but its strong reduction with the rotational disorder. More precisely, in the case of distribution disorder, the investigation revealed a significant reduction in both electronic conductance and electronic thermal conductance when compared to the ideal periodic structure. However, due to the qualitatively unchanged band gap, the Seebeck coefficient exhibited a notably lesser reduction. As a consequence, these factors lead to a significant decrease in the PF, primarily due to the substantial reduction in conductance, rather than a band gap closing as showed in the models treated in section 3.1, highlighting the importance of precise control over the positioning of BN rings when constructing BNC materials for thermoelectric purposes. Similarly, the findings concerning rotational disorder showed a decline in electronic and electronic thermal conductance when increasing the disorder. Yet, unlike distribution disorder, the overall conductance reduction was marginally less pronounced. As anticipated by DFT calculations for periodic Models 3 and 4, the introduction of rotational disorder leads to a gradual chiral symmetry restoration, resulting in a progressive narrowing of the band gap. Consequently, a decrease in the Seebeck coefficient with the rotational disorder concentration is reported. Nevertheless, merging these observations revealed that the PF exhibited comparable values to those computed for scenarios involving distribution disorder. Finally, these results underscore the importance of controlling the position of the BN ring to maintain the conductivity as high as possible, even if the Seebeck coefficient would not be affected by it. On the other side, the rotational disorder would be one of the key factors in tuning the thermoelectric properties of BNC materials, as already predicted in DFT calculations, due to its strong influence on the electronic properties.

\subsection{Thermoelectric efficiency}

\begin{figure}[!ht]
    \centering
    \includegraphics[width=12cm]{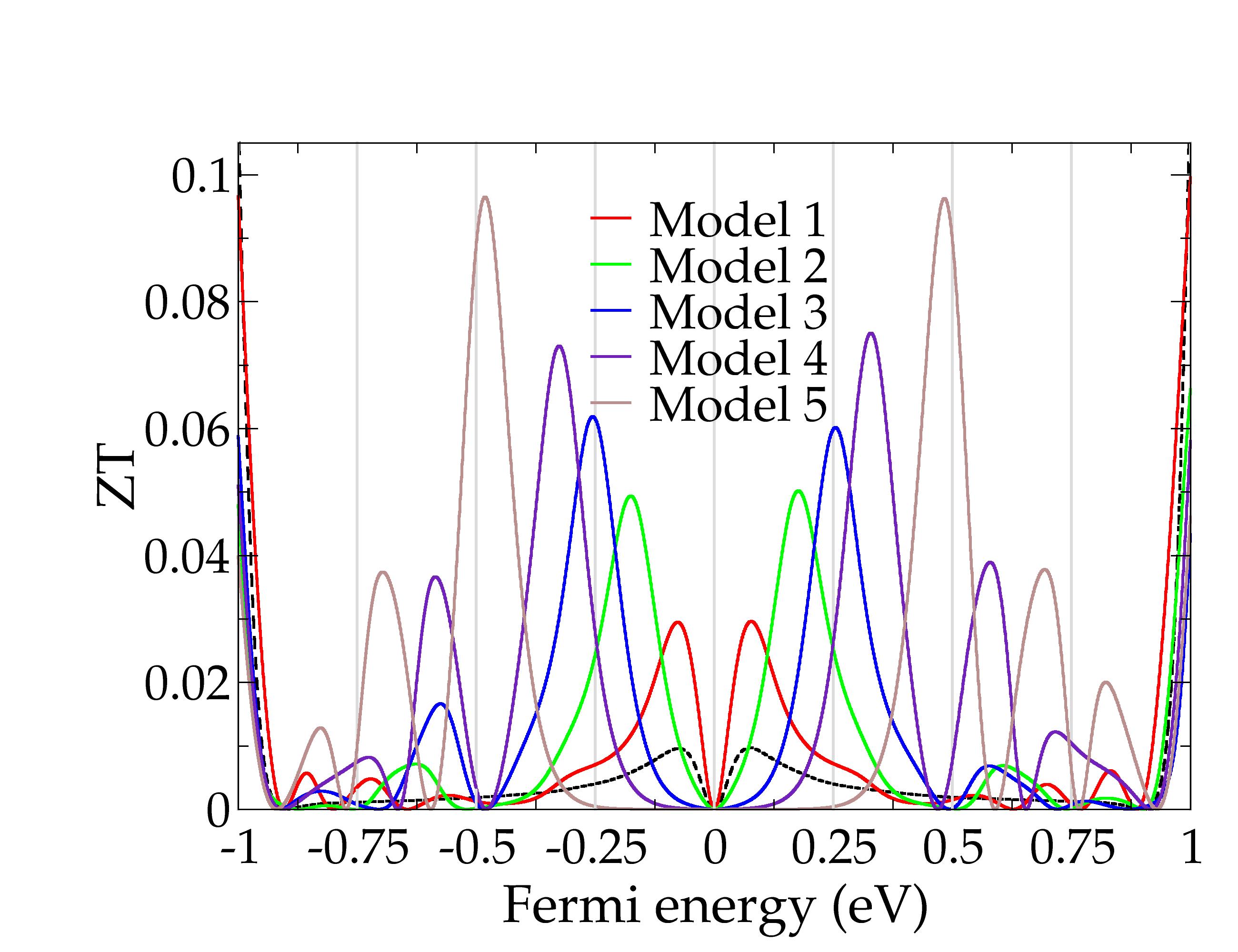}
    \caption{Thermoelectric figure of merit ZT computed for BN-ring doped graphene systems with all parallel rings as considered in Fig.\ref{fig:upup}. ZT of pristine graphene is reported in dashed black line.}
    \label{fig:ZT}
\end{figure}

Upon a thorough investigation into the influence of BN-doping on electron thermoelectric properties, another crucial aspect to consider is lattice thermal conductance. In fact, pristine graphene exhibits an exceptionally high lattice thermal conductivity \cite{Balandin2008} attributed to the remarkable strength of its C-C sp$^2$ bonds and the uniformity of its lattice structure.  As a result, the thermoelectric figure of merit of graphene falls below 0.01, that is a relatively low efficiency. For this reason, it is of crucial importance to assess the possibility of reducing the $K_l$ when doping graphene with borazine rings. In particular,  the latter plays a significant role in tuning the values of the thermoelectric figure of merit ZT. Similarly to what has been done in the study of the electronic part, the $K_l$ for each of the periodic models has been estimated and reported in Table \ref{tab:Kl}.
\begin{table}[]
    \centering
\begin{tabular}{|c|c|c|c|c|}
\hline
K$_l$(W m$^{-1}$K$^{-1}$) & 0\%  & 17\% & 33\% & 50\% \\
\hline
Graphene & 1.31 &      &      &      \\
Model 1  & 1.08 & -    & -    & -    \\
Model 2  & 0.96 & -    & -    & 0.98 \\
Model 3  & 0.88 & -    & 0.88 & -    \\
Model 4  & 0.86 & -    & 0.86  & 0.86 \\
Model 5  & 0.80 & 0.80 & 0.79 & 0.81 \\
\hline
\end{tabular}

\caption{Lattice thermal conductance (in W K$^{-1}$m$^{-1}$) calculated with the 4NN FC model for pristine graphene and BN-ring doped systems with Models 1, 2, 3, 4 and 5 (see Fig.\ref{fig:upup}). For each BN-ring doped system, possible anti-parallel (0$\%$, 17$\%$, 33$\%$ or 50$\%$) configurations are considered.}
\label{tab:Kl}
\end{table}
$K_l$ is found to be notably independent on the variation of the number of rotated BN rings. This insensitivity can be easily attributed to the fact that rotational manipulation of BN-rings significantly alters the electronic band structure through chiral symmetry breaking while having relatively minor effects on the phonon dispersion. As a consequence, $K_l$ exhibits discrepancies of no more than about 0.02 W m$^{-1}$K$^{-1}$ for a given BN-ring concentration but with different percentages of anti-parallel rings. Moreover, a significant decrease of $K_l$ with the increase of BN concentration is observed, with the largest reduction of $K_l$ by 40\% for Model 5.
By merging the $K_l$ values with the previously explored thermoelectric properties, the figure of merit can thus be estimated. In Fig. \ref{fig:ZT}, the ZT values for all the previously discussed models are reported in a parallel orientation. The results obtained for different anti-parallel configurations of these models are also presented in section 4 of Supplementary Material.

As expected, a trend similar to the one of the PF is found. The maximum ZT reaches $\sim$0.1, achieved within Model 5. Even though this is not a large value, it is a significant improvement, i.e., about ten times larger than the value in pristine graphene. The obtained results are attributed to the limitation of the considered technique in achieving a sufficiently small K$_l$ compared to alternative approaches, such as cross-plane techniques \cite{Hung2014,Sadeghi2017,Olaya2017,Olaya2019}, oxygen functionalization \cite{Zhang2014,Mu2014} or controlled isotope doping \cite{Anno2014}. Combining the engineering approach considered here with other nanostructuring methodologies could further reduce K$_l$ and thus lead to a larger enhancement of thermoelectric efficiency.

\section{Conclusions}
This work explores how introducing BN rings into 2D graphene impacts its thermoelectric properties. The incorporation of BN-ring is found to consistently enhance the thermoelectric properties of pristine graphene. Aligning BN-rings in a parallel orientation breaks the sublattice chiral symmetry, resulting in the formation of a band gap and a notable increase in the Seebeck coefficient, leading to the efficient separation of electrons and holes contribution in BN-doped graphene. Indeed, the incorporation of BN rings with an anti-parallel orientation offers a means to accurately adjust thermoelectric properties by concurrently manipulating both the rotational disorder and the concentration of BN rings. By precisely tuning these factors, it is possible to achieve a power factor comparable to the respective values in the parallel-oriented scenario. Moreover, weak disorder in the position and orientation of BN-rings in aperiodic cases shows a drastic decrease in the conductance, leading to a reduction in the PF. Lastly, the lattice thermal conductivity shows a reduction of up to 40$\%$ in high BN-ring concentration cases, with a qualitative insensitivity to BN-ring rotational disorder. By combining these results, a maximal ZT value of $\sim$0.1 is achieved.  In summary,  the incorporation of BN-co-doping allows for tuning of the thermoelectric properties in 2D graphene. This not only provides a means to improve the PF, paving the way for potentially increased output power density in thermoelectric devices, but also presents a pathway for precise modulation of each thermoelectric aspect by adjusting various doping parameters, including concentration and ring orientation.

\section*{Acknowledgments}
The authors acknowledge fundings from the European Union's Horizon 2020 Research and Innovation programme under the Marie Sklodowska-Curie entitled "STiBNite" (N$^{\circ}$ 956923), from the Flag-Era JTC project "MINERVA" (N$^{\circ}$ R.8006.21), from the Pathfinder Project "FLATS" (N$^{\circ}$ 101099139), from the F\'ed\'eration Wallonie-Bruxelles through the ARC project "DREAMS" (N$^{\circ}$ 21/26-116),  from the EOS project "CONNECT" (N$^{\circ}$ 40007563) and from the Belgium F.R.S.-FNRS through the research project (N$^{\circ}$ T.029.22F). Computational resources have been provided by the CISM supercomputing facilities of UCLouvain and the C\'ECI consortium funded by F.R.S.-FNRS of Belgium (N$^{\circ}$ 2.5020.11).


\clearpage

\begin{center} \Huge
\end{center}
\begin{center} \Huge
\end{center}

\begin{center}
\huge \textbf{\underline{Supplementary Material}}
\end{center}
\setcounter{section}{0}
\setcounter{figure}{0}
\renewcommand{\thefigure}{S\arabic{figure}}

\begin{center} \Huge
\end{center}
\begin{center} \Huge
\end{center}

\section{Different configurations of ring position in DFT calculations}

\begin{figure}[!h]
    \centering
    \includegraphics[width=15.1cm]{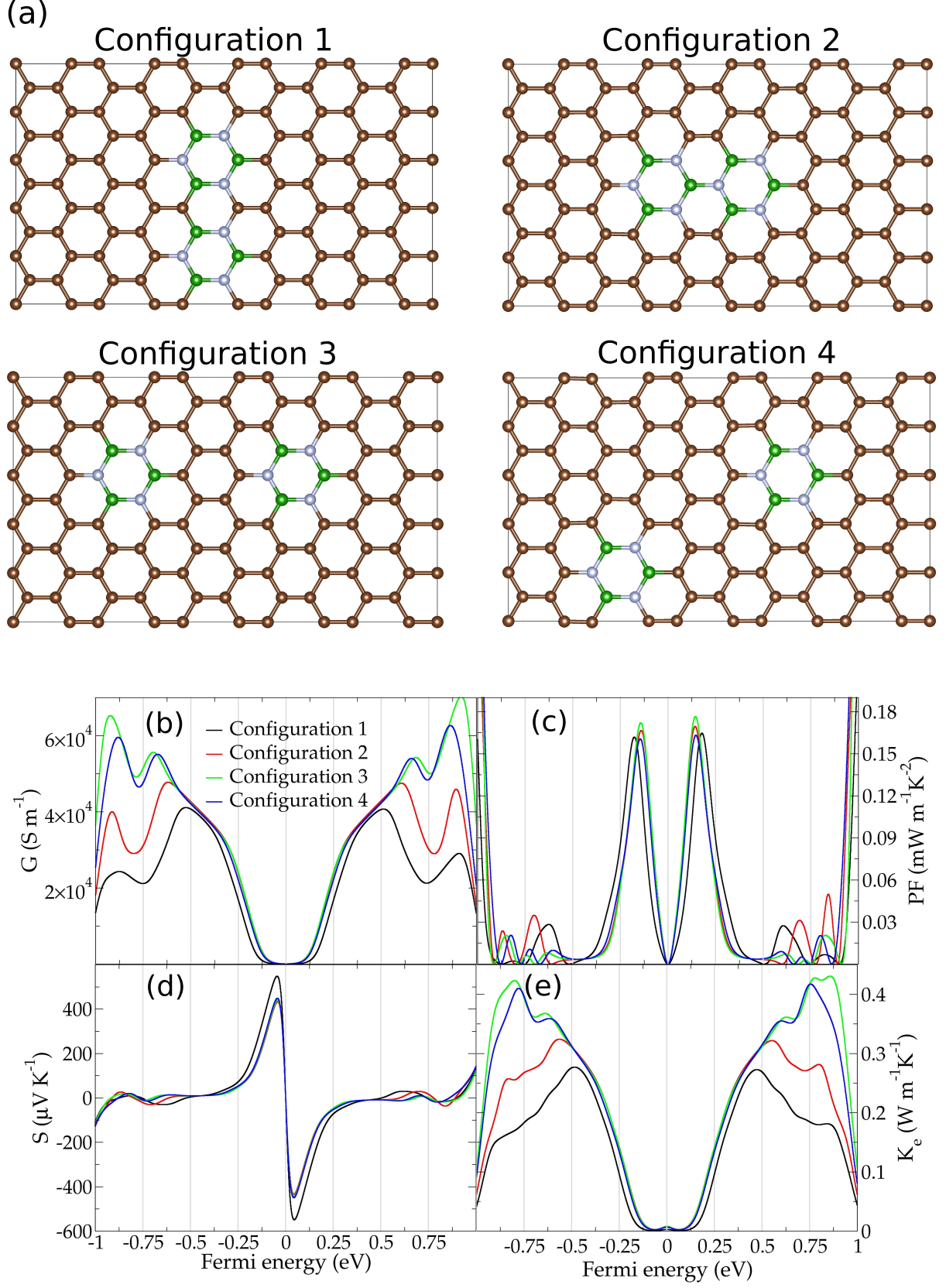}
    \caption{(a) Atomic structures of Model 2 with four different doping distributions (configurations 1,2,3 and 4) and their corresponding (b) electronic conductance (\textit{G}), (c) Seebeck coefficient (\textit{S}), (d) power factor (\textit{PF}), (e) electronic thermal conductance ($K_e$) as a function of the Fermi level.}
    \label{fig:model2_conf}
\end{figure}

When there are more than 2 rings in the DFT unit cell, changing the position of BN-rings can form different doping configurations. To illustrate the dependence of the thermoelectric properties on these configurations, various doping topologies have been examined and presented here for Model 2. In particular, four different doping configurations and their corresponding thermoelectric properties are shown in Fig. \ref{fig:model2_conf}. While large variations in conductance and thermal conductance are observed only at high energies, the band gap remains qualitatively similar in all cases, resulting in comparable Seebeck coefficients. Consequently, the combination of these factors leads to a qualitatively similar power factor. All these results demonstrate the negligible impact of the position of the BN-rings within the unit cell on the thermoelectric properties. As a result, a single doping configuration for each concentration, as considered in the main text (Fig. 2), can be considered as a representative model for all the calculations.

\newpage

\section{Different configurations of anti-parallel BN rings in DFT calculations}

\begin{figure}[!h]
    \centering
    \hspace*{-0.8cm}
    \includegraphics[width=15.7cm]{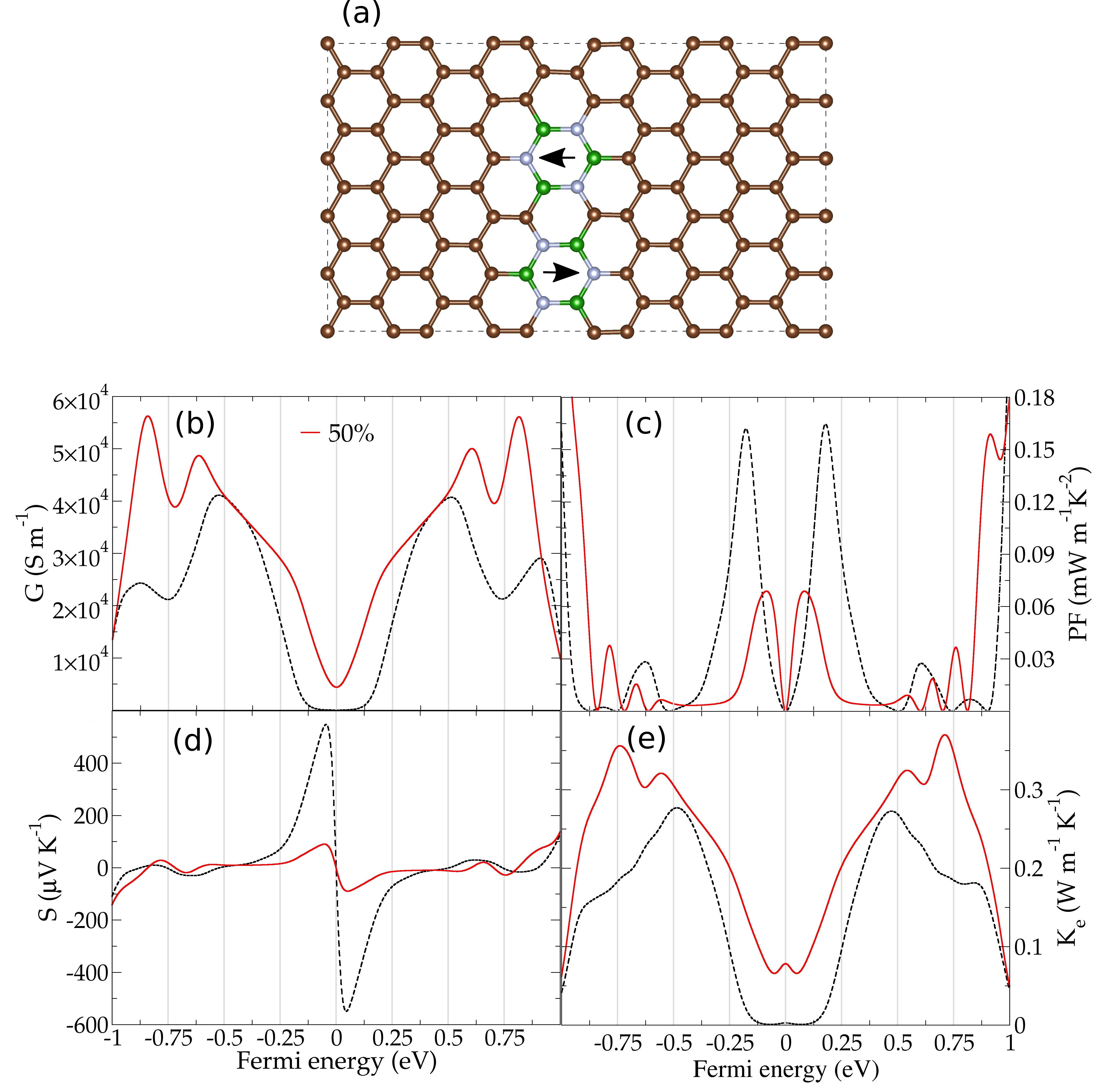}
    \caption{(a) Atomic structures of Model 2 with anti-parallel BN rings and its corresponding (b) electronic conductance, (c) Seebeck coefficient, (d) power factor, (e) electronic thermal conductance as a function of the Fermi level. The results for the same Model 2 with all parallel rings (i.e., black dashed lines) are added for comparison.}
    \label{fig:model2}
\end{figure}

\begin{figure}[!h]
    \centering
    \includegraphics[width=15.7cm]{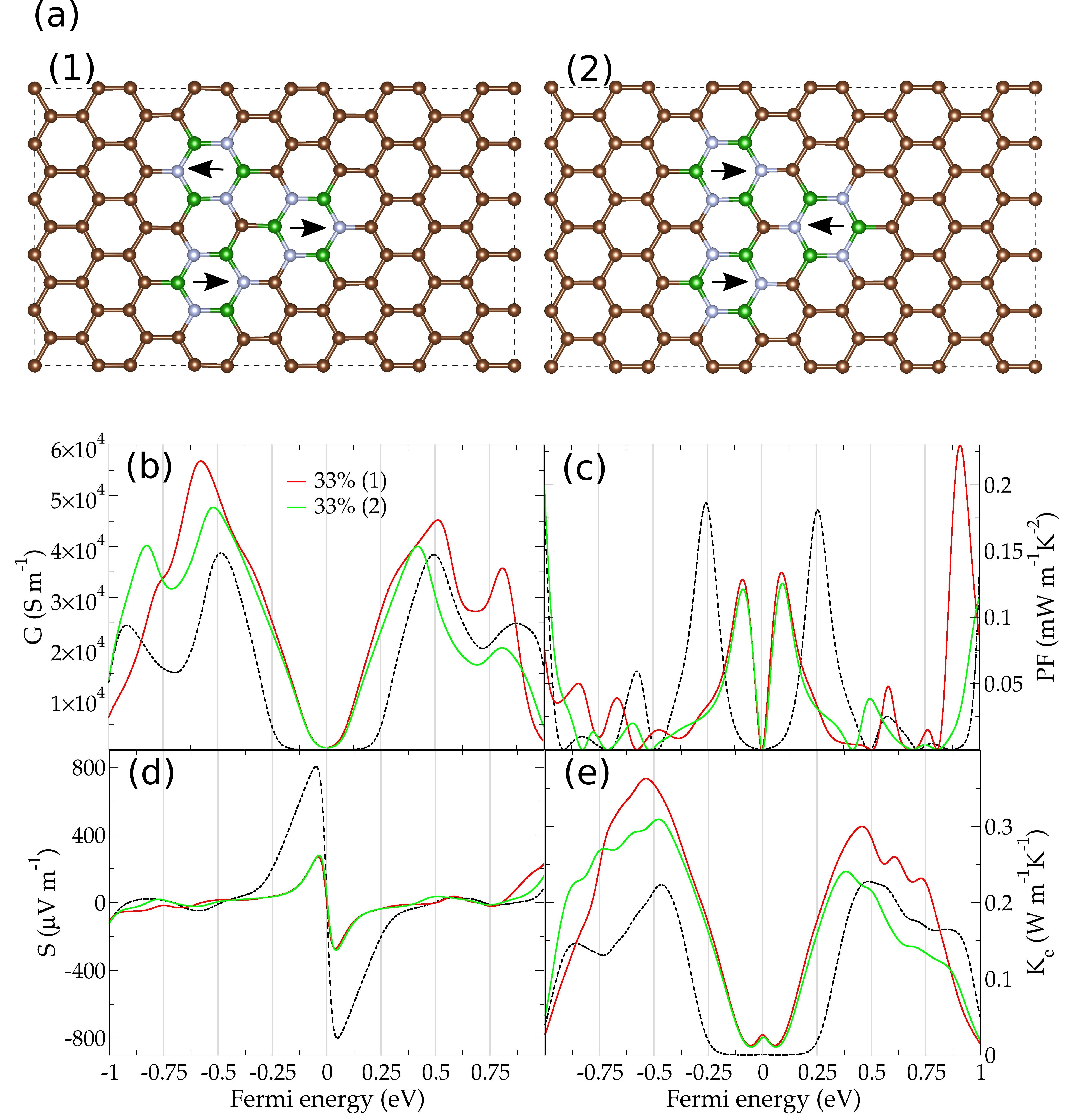}
    \caption{(a) Atomic structures of Model 3 with two anti-parallel configurations (namely (1) and (2) related to the 33$\%$ anti-parallel ring cases) and their corresponding (b) electronic conductance, (c) Seebeck coefficient, (d) power factor, (e) electronic thermal conductance as a function of the Fermi level. The results for the same Model 3 with all parallel BN-rings (i.e., black dashed lines) are added for comparison.}
    \label{fig:model3}
\end{figure}

\begin{figure}[!ht]
    \centering
    \hspace*{-0.8cm} 
    \includegraphics[width=15.7cm]{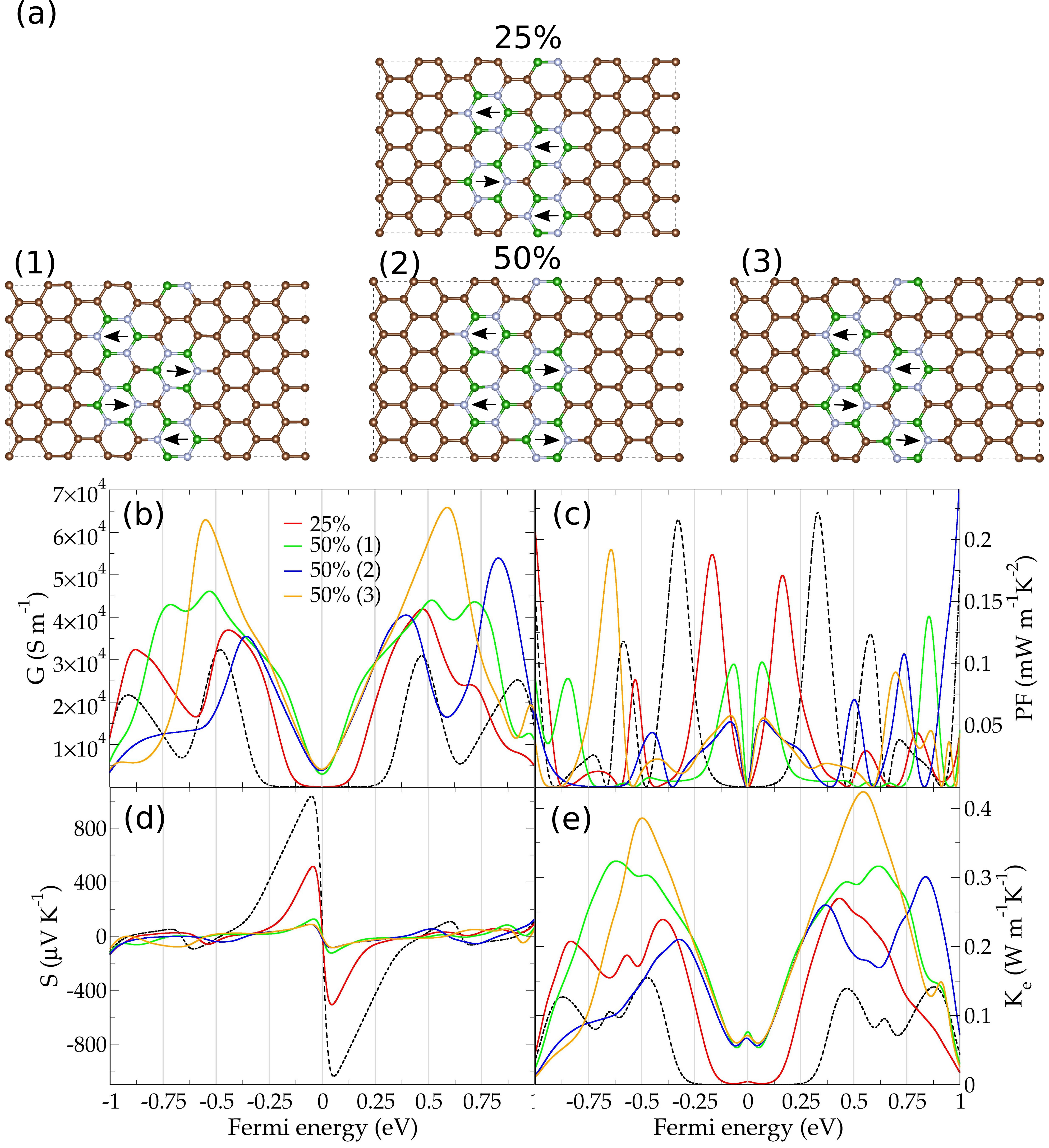}
    \caption{(a) Atomic structures of Model 4 with one 25\% anti-parallel ring configuration and three 50\% anti-parallel ring configurations (namely, (1),(2) and (3)). Correspondingly, their (b) electronic conductance, (c) Seebeck coefficient, (d) power factor, (e) electronic thermal conductance as a function of the Fermi level. The results for the same Model 4 with all parallel BN-rings (i.e., black dashed lines) are added for comparison.}
    \label{fig:thermo_4_si}
\end{figure}

\begin{figure}[!h]
    \centering
    \includegraphics[width=15.7cm]{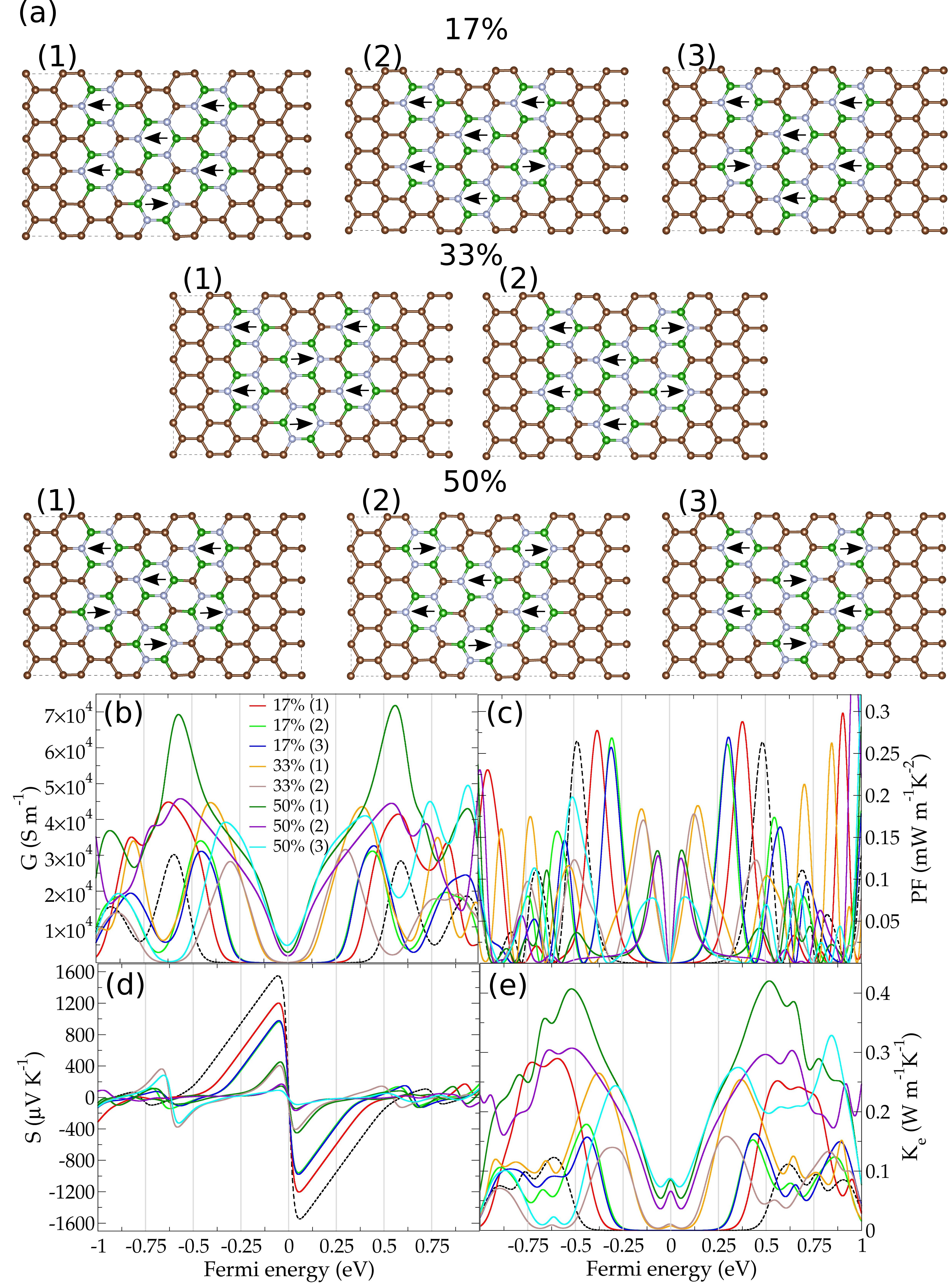}
    \caption{(a) Atomic structures of Model 5 with different anti-parallel configurations, including 17\%, 33\% and 50\% anti-parallel rings. Correspondingly, their (b) electronic conductance, (c) Seebeck coefficient, (d) power factor, (e) electronic thermal conductance as a function of the Fermi level. The results for the same Model 5 with all parallel BN-rings (i.e., black dashed lines) are added for comparison.}
    \label{fig:model6_si}
\end{figure}

The impact of BN-ring orientations on the thermoelectric properties is considered herewith for different concentrations (i.e., models 2, 3, 4, and 5 in the main text). For each model, different percentages of anti-parallel BN rings are calculated and presented (see Figures \ref{fig:model2}, \ref{fig:model3}, \ref{fig:thermo_4_si}, and \ref{fig:model6_si}). Overall, the thermoelectric effects are shown to be significantly affected (i.e., weakened) when increasing the percentage of anti-parallel rings. The essential reason for these results is the reduction of bandgap when partially rotating BN rings as discussed in the main text as well as in Ref.\cite{Caputo2022}. On the contrary, similar to the results discussed in Fig. \ref{fig:model2_conf}, the obtained results are not strongly dependent on the position of rotated rings.
\newpage

\clearpage

\section{Tight-Binding Calculations}

The electronic transport in large aperiodic BNC systems is computed using a $\textit{p}_z$-orbital tight-binding Hamiltonian \cite{Fiori2012S,Nguyen2012S} with a proper adjustment for B-N hopping energies. The validity of this TB Hamiltonian has been demonstrated by comparing its computed band structures with corresponding DFT results (see in Ref. \cite{Caputo2022S}).

\section{Thermoelectric efficiency}

\begin{figure}[!h]
    \centering
    \hspace*{-0.8cm} 
    \includegraphics[width=15.7cm]{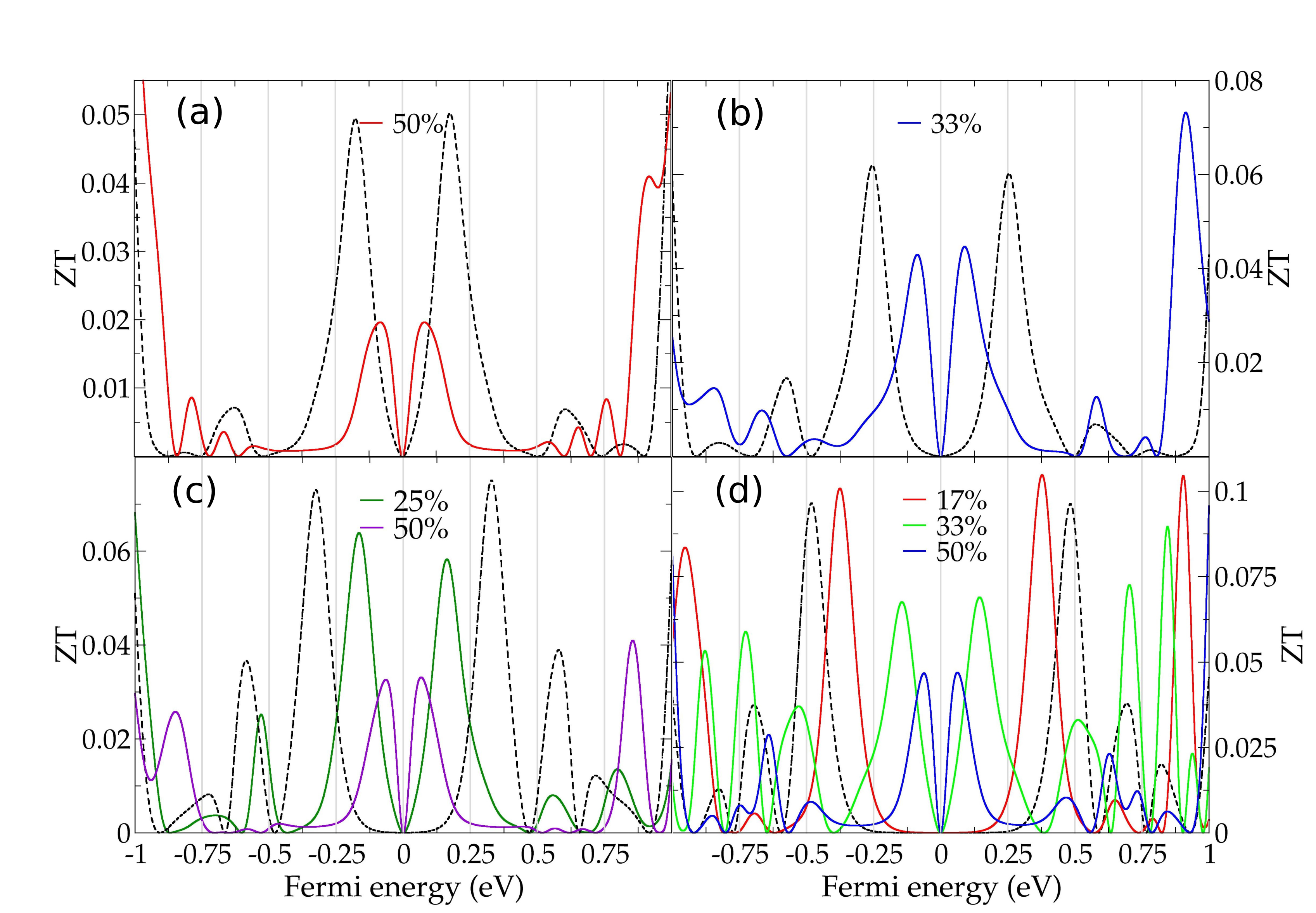}
    \caption{Thermoelectric figure of merit ZT for (a) Model 2 with 50$\%$ anti-parallel rings, (b) Model 3 with 33$\%$ anti-parallel rings, (c) Model 4 with 25$\%$ and 50$\%$ anti-parallel rings, (d) Model 5 with 17$\%$, 33$\%$ and 50$\%$ anti-parallel rings. The black dashed lines present the results obtained for the case of all parallel rings.}
    \label{fig:ZT_updown}
\end{figure}

The ZT values for different anti-parallel configurations are reported in Fig. \ref{fig:ZT_updown}. These configurations are formed by partially rotating BN rings in the DFT unit cells, i.e., Models 2, 3, 4 and 5 presented in Fig.\ref{fig:upup}. These presented results are obtained for the corresponding periodic systems by DFT calculations computing the electronic transport and force-constant models computing the lattice thermal conductance. 

\end{document}